\documentclass[sigconf, 10pt]{acmart}
\def\BibTeX{{\rm B\kern-.05em{\sc i\kern-.025em b}\kern-.08emT\kern-.1667em\lower.7ex\hbox{E}\kern-.125emX}}

\usepackage{enumitem}
\usepackage{url}
\PassOptionsToPackage{hyphens}{url}

\usepackage{color}
\usepackage{listings}

\lstdefinelanguage{JavaScript}{
	keywords={typeof, new, true, false, catch, function, return, null, catch, switch, var, if, in, while, do, 
	else, case, break},
	keywordstyle=\color{blue}\bfseries,
	ndkeywords={class, export, boolean, throw, implements, import, this},
	ndkeywordstyle=\color{darkgray}\bfseries,
	identifierstyle=\color{black},
	sensitive=false,
	comment=[l]{//},
	morecomment=[s]{/*}{*/},
	commentstyle=\color{blue}\ttfamily,
	stringstyle=\color{blue}\ttfamily,
	morestring=[b]',
	morestring=[b]"
}
\usepackage{tikz} 
\newcommand*\circled[1]{\tikz[baseline=(char.base)]{
		\node[shape=circle,draw,inner sep=0.5pt] (char) {#1};}}
	
\copyrightyear{2021}
\acmYear{2021}

\acmConference[ZHAW Digitcalcollection - Technical 
Report ]{Technical Report}{March 2021}{Zurich}

\renewcommand\footnotetextcopyrightpermission[1]{} 

\usepackage{orcidlink}

\begin{document}
\title{Web Content Signing with Service Workers}

\author{Thomas Sutter}
\authornote{Main contributor}
\orcid{0000-0003-2649-3299}
\email{suth@zhaw.ch}
\author{Peter Berlich}
\email{berp@zhaw.ch}
\author{Marc Rennhard}
\email{rema@zhaw.ch}
\affiliation{%
	\institution{Zurich University of Applied Sciences - Institute of Applied 
		Information Technology}
	\streetaddress{Obere Kirchgasse 2}
	\city{Winterthur}
	\state{Zurich}
	\country{Switzerland}
	\postcode{8401}
}
\author{Kevin Lapagna}
\email{klapagna@redhat.com}
\authornote{Main contributor}
\affiliation{%
	\institution{}
	\streetaddress{XXX}
	\city{Zurich}
	\state{}
	\country{Switzerland}
	\postcode{8003}
}

\author{Fabio Germann}
\email{fabio.germann@dswiss.com}
\affiliation{%
	\institution{DSwiss AG}
	\streetaddress{Badenerstrasse 329}
	\city{Zurich}
	\state{}
	\country{Switzerland}
	\postcode{8003}
}

\renewcommand{\shortauthors}{T. Sutter, et al.}
\begin{abstract}
Securing the communication between a web server and a browser is a fundamental task of securing the 
World Wide Web. Websites today rely heavily on HTTPS to set up secure connections. In recent years, 
several incidents undermined this trust and therefore the security of the HTTPS system. In this 
paper we introduce an approach allowing to secure JavaScript files in case a HTTPS connection 
between web server and browser is compromised. Our paper presents a solution to safeguard the 
user's browser so that it only processes content (e.g., JavaScript or HTML) that was 
genuinely provided by the web application service providers themselves. Our solution makes use of 
service workers, a recently proposed W3C Candidate Recommendation enabling applications to take 
advantage of persistent background processing, including hooks to enable bootstrapping of web 
applications while offline. It demonstrates how service workers are able to validate the integrity of 
JavaScript files within the client's browser and how service workers are used to detect and mitigate 
malicious JavaScript files.
\end{abstract}
\maketitle
%

\section{Introduction}\label{introduction}
Transport Layer Security (TLS) allows to encrypt and secure data traffic between two parties in a 
computer network. It is widely used in the World Wide Web (WWW) as part of the Hyper Text Transfer 
Protocol Secure (HTTPS) to protect traffic between web browsers and web servers. It is touted as one 
of the most common data transmission security mechanism used today, and its newest version 1.3 is 
considered state-of-the-art in digital communication protection. But even though TLS is considered 
highly secure in regard to its underlying cryptographic primitives and protocols, it has one notable 
weakness in the way it is applied in common web browsers: Its dependency on a Public Key 
Infrastructure (PKI) based on Certificate Authorities (CAs).

Browsers and operating systems come with a prepackaged set of trusted CA certificates. When a web 
browser instantiates a TLS connection to a web server, it has to check whether the X.509 certificate 
issued to and provided by the web server is valid. One important step to check the validity of the 
certificate is checking whether it was signed by one of the prepackaged trusted CAs (if 
not removed by the user).

For instance, the Mozilla Firefox browser comes with a prepackaged list of about 150 trusted Root 
CAs~\cite{mozilla_ca_certificate_list} that in turn can deploy and authorize intermediate CAs, which in 
the end will sign the certificates used for TLS. The weakness in this system comes from the fact, that if 
\textit{any} of the 150 Root CAs or \textit{any} of the many more intermediate CAs gets compromised, 
an attacker might be able to forge a valid certificate for an arbitrary chosen domain. 

If this happens, a man-in-the-middle attack becomes feasible for the holder of the forged certificate 
and as 
consequence, all transferred data can be read and manipulated by an attacker at will. 

In the case of a web application, this even means that an attacker can hijack an ongoing user session 
and execute 
arbitrarily operations within the web application in the user's name. Depending on the web application 
this may for instance lead to financial or reputational loss for the service provider and/or user. As 
provider or user, one therefore has to not only to trust in the proper operation of modern 
cryptography (i.e., a working TLS connection) but also in the capabilities and security of hundreds of 
companies, their IT systems and thousands of their employees 
all around the world.

Multiple well-documented incidents where attackers were able to gain illegitimate 
certificates have already occurred~\cite{certificate_authority_incidents, berkowsky2017security}. One of 
the most notorious incidents involved the CA DigiNotar, which issued a fraudulent wildcard certificate 
for Google 
that was later used in a man-in-the-middle attack~\cite{prins2011diginotar}. Additionally, surveillance 
disclosures, brought into motion by ex-NSA contractor Edward Snowden in 2013, indicate that certain 
organizations might be able to break or circumvent TLS in yet undisclosed ways~\cite{nsa_flying_pig}. 
Sloppy CA processes, malicious actors, installed weak root certificates 
(e.g.,~\cite{superfish},~\cite{kaspersky_mim}), TLS stripping, or just careless users ignoring browser 
warnings for questionable certificates, can as well lead to successful man-in-the-middle attacks.

In this paper, we present a new approach to address such attacks with the help of a component 
available in virtually all modern browser by default--- service workers. We show how it can be 
leveraged to set up an additional authenticated encryption layer on top of HTTPS which can be 
implemented in plain JavaScript. The fundamental challenge to be addressed here is the delivery 
mechanisms of the JavaScript code, as all JavaScript code that is run on the client's browser is 
delivered over a 
HTTPS connection (chicken-and-egg-problem).

This leads to the problem, that a man-in-the-middle could modify the JavaScript files in transit and 
disable the JavaScript encryption. As consequence, any additional layer of encryption is only effective 
if it is tamper resistant against modifications by a man-in-the-middle.

This example highlights that there are two problems which need to be solved in order to establish a 
trusted layer of encryption in JavaScript over a HTTPS connection that might have been intercepted 
using a rogue certificate:
\begin{enumerate}
	\item Exchanging encryption keys over an insecure HTTPS connection, assuming that an attacker is 
	listening to the communication.
	\item Delivering JavaScript that is cryptographically proven to not have been modified by a 
	man-in-the-middle.
\end{enumerate}
In general, these problems can be tackled in two different ways. We can either distribute a trusted 
component within the browser over another channel, which is believed to be resistant to 
man-in-the-middle attack, or we can set up a trust on first use (TOFU) system. 

Our solution proposes a TOFU system which allows to solve the mentioned problems by using service 
workers. No 
user interaction is required to get this solution up and running, since service workers are widely 
supported by most browsers today \cite{caniuse_service_workers} and used in web sites all over the 
world. We will show that service workers, once installed, can be used to prevent man-in-the-middle 
attacks from powerful adversaries with rogue certificates.

The remainder of this paper is structured as follows. Section 2 discusses alternative approaches to 
solving the problem. In Section 3, we provide more details on the attack scenario considered in this 
paper. Section 4 and 5 introduce the basic principles of service workers and how we use them as 
security measure. Section 6 describe a implementation example and shows how incidents can be 
handle in practise.

The setup and evaluation methodology for our evaluation is presented in Section 7 and the 
corresponding results in Section 8. The paper is concluded with a discussion of limitations in Section 9 
and an outlook in Section 10.

\section{Related Work}\label{related_work}
The shortcomings in CA PKI lead to the situation that even with two uncompromised endpoints, 
there might be the possibility of a man-in-the-middle attack tapping into the connection. Various 
efforts to counter this kind of attacks on HTTPS have been conducted in the past, resulting in 
mechanisms and protocols such as HTTP Public Key Pinning (HPKP) ~\cite{httpKeyPinningPaper}, 
DNS-based Authentication of Named Entities (DANE) ~\cite{dane} or Certificate Transparency (CT) 
~\cite{certificateTransparency}. HPKP is considered deprecated since version 69 of Chrome due to 
having dangerous side effects ~\cite{google_remove_hpkp} and DANE has no major browser support. 
At the time of writing only CT is considered a state-of-the-art mechanism and it is widely deployed. 

Certificate Transparency ensures that issued certificates by CAs are logged and can be monitored and 
audited. In order to ensure that no rogue certificates are issued, CAs and domain owners 
have to monitor the CT logs continuously. In short, with CT it is possible to 
retrospectively detect that a rogue certificate was issued, but CT alone does not prevent the attack. 

Further actions by the CA and the domain owner have to be taken in order to mitigate the attack. The 
CA and domain owners have to detect the attack and revoke the rogue certificate. The revocation can 
take some time which may allow the attacker to do financial or reputational damage.
\section{Attack Scenarios}\label{attack_scenarios}
 The main attack scenario against which we evaluated our approach was a man-in-the-middle attack on 
 TLS with a certificate accepted by the browser. 

We further distinguish between passive and active man-in-the-middle attacks. In a \textit{passive} 
man-in-the-middle attack, the attacker is able to read all transferred data in cleartext without 
interfering with the traffic. In an \textit{active} man-in-the-middle attack, the attacker is additionally 
able to 
inject and/or manipulate any transmitted content. Both kinds, active and passive, are assumed 
to be undetectable by the browser or the web server (in absence of additional countermeasures).

Starting from the premise that the cryptographic primitives of TLS themselves withstand the 
attempt to accomplish a successful man-in-the-middle attack, it is required for the attacker to be in 
possession of a private key, matching a X.509 certificate that the victims browser accepts. This can 
happen in the following scenarios:
\begin{itemize}
	\item A genuine CA issues a certificate for the concerning domain (by either cheating 
	or forcing (e.g., as state actor) the CA to comply).
	
	\item The attacker steals the private key for a valid CA certificate (e.g., by exploiting a weakness in 
	the CA's IT infrastructure).
	
	\item The attacker steals the private key for a genuine certificate of the concerning domain (e.g., by 
	exploiting a weakness in a web server).
	
	\item The attacker exploits an installed weak non-standard CA certificate (installed, e.g., due to 
	company-policy, anti-virus software, malware, etc.).
\end{itemize}

On a technical level it is irrelevant which of these scenarios happens, in 
every case, the connection has to be considered insecure i.e., privacy, authenticity and integrity 
provided by TLS, can not be taken for granted.

Additionally, the timing of an attack is considered. We define three possible timing 
scenarios:
\begin{enumerate}[label=(\Alph*)]
	\item An attack is conducted from the beginning of the TLS handshake. In this scenario the user has 
	never visited the domain before and an attacker can tamper the connection from the first moment in 
	time (from the first byte) when the browser attempts a connection to the domain.
	\item The client has already visited a domain once (without a man-in-the-middle attack). The attacker 
	can tamper the connection from the first moment in time of the second connection to the same 
	domain.
	\item In the third scenario, the TLS connection between the endpoints have already been established 
	and the domain has been visited more than once by the client. An attacker modifies or intercepts 
	content during an active TLS connection.
\end{enumerate}

\section{Service Workers}\label{fundamentals}
\begin{figure*}[t]
	\centerline{\includegraphics[width=0.75\textwidth]{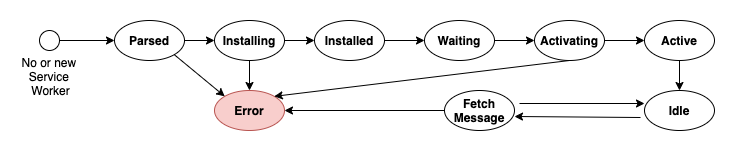}}
	\caption{Service workers state diagram.}
	\label{service_worker_life_cycle}
\end{figure*}
Service workers are part of the HTML Living Standard~\cite{html_living_standard}. They are specified in 
a W3C Editor's Draft~\cite{service_workers_editors_draft} which is under 
development at the time of writing. Service workers were introduced in Chrome and Mozilla Firefox in 
2015, with Safari 
following in 2018. They are now supported and enabled by default in the browsers of 
around 90\% of all desktop users. The availability in browsers for mobile users is even 
higher~\cite{caniuse_service_workers}.

Similar to browser plugins, service workers can act as a proxy-like entity between the web server and 
the regular runtime of the user`s browser. In contrast to browser plugins, service workers do not need 
to be installed manually by the user. Originally they were introduced as a way to provide a better 
user experience when in case of temporary network outages. In such scenario, the service worker 
can answer the browser requests with cached or self-generated content, instead of an actual web 
server response. With this, it is for example possible to fully operate a web-based e-mail client when 
offline. In this case the user can search, read and even answer e-mails that were 
cached within a service worker in advance. Pending in- and outgoing messages can be processed at a 
later point in time when the network is available again.

Understanding the life cycle of service workers is a crucial precondition when implementing a security 
solution utilizing them. In the following paragraphs, we first present the basic life cycle of a service 
worker given by W3C Editor's Draft~\cite{service_workers_editors_draft} and then take a closer look 
at some important corner cases that are relevant for our approach.

\textbf{Parsed:} Before we can make use of a service worker, it has to be registered, installed 
and activated. A service worker is always registered for a certain scope. The scope can either be the 
whole domain (e.g., \url{https://example.com/}) or a sub-path (e.g., \url{https://example.com/sub/}). 
Only one service worker can be active at any given time. If a second service worker is registered or a 
new version is made available, the current one is halted and replaced. The individual steps take place as 
follows:

\textbf{Installing:} The registration process can be initiated with the following code in one of the 
regular loaded JavaScripts: \textit{ navigator.serviceWorker.register(`sw.js')}. The browser will then 
request the service worker (defined in this example in \textit{sw.js}) from the web server, whereupon it 
will be registered, installed and cached (as described in the following points). This process will only 
succeed if the service worker is served over TLS, otherwise the installation is rejected by the browser 
as stated in~\cite{service_workers_editors_draft}. Serving the service worker over TLS is a security 
measure enforced by the browser to prevent the installation of a tampered service worker through an 
insecure connection. As an important limitation, at this point, we still have to trust that TLS will work 
properly, and that no active man-in-the-middle attack is ongoing. This means the installation of a 
service worker relies on the fact that at installation time no active man-in-the-middle attack is 
conducted. This trust on first use problem is further discussed in Section ~\ref{tofu_discussion}.

Once the browser receives the service worker, a separate thread will start in the browser that will 
execute it. During this phase an \textit{install} event can spawn functions to perform arbitrary tasks, 
like pre-caching of static web application assets.

From the moment the service worker is installed, it can be seen as a kind of short-lived standalone 
extension within the browser. Most importantly, it has its own states and lifetime which are 
independent of the displayed browser window. After a successful installation, the service worker 
changes from the \textit{installing} into the \textit{active} state (see 
Figure~\ref{service_worker_life_cycle}). An \textit{activate} event is fired at this point. Note that the 
service worker still has to claim the web page.

\textbf{Waiting:} After a service worker is installed, it does \textit{not} get active immediately. In case 
that another service worker is already installed it goes into the waiting state until the old service worker 
can be replaced. This happens either when the page is revisited by the user or when the new service 
worker uses \textit{clients.claim()}. With \textit{clients.claim()}, the service worker can skip the waiting 
time using the \textit{skipWaiting()} method. In this case, the service worker replaces the old one 
immediately after its install routine returns successfully.

\textbf{Active:} In this state the service worker is installed and active, but does not operate as one 
might expect; it does not intercept any requests yet. For consistency reasons, a service worker usually 
only reacts to requests originating from web pages which were loaded through the respective service 
worker. This means all windows of that website must be closed and the users has to revisit it again at a 
later point in time.\footnote{A force reload to ignore cached content will yield the same result.} On this 
second visit, the service worker claims the page even \textit{before} the first web request to the 
domain is made. This means from this point on it can intercept all requests/responses to and from the 
website. 

The service worker can react to the \textit{activate} event and can then call \textit{clients.claim()}. This 
will lead to the desired situation where the service worker gets control over submitted requests and 
responses. Note that all of this happens asynchronously, so there are no guarantees when exactly the 
page gets claimed by the service worker.

\textbf{Idle:} Once fully active, the service worker can listen to requests in the form of \textit{fetch}
events coming from the respective web page~\cite{service_workers_editors_draft}. It can take full 
control of a request by calling \textit{event.respondWith()} on the respective event. The service worker 
can now forward, manipulate or dismiss the request. When it receives a response from the web server, 
it can once again freely forward, manipulate or dismiss this response. It therefore acts now like a proxy 
server in between the regular part of the browser and the website.

\textbf{Updating}: From time to time a service worker has to be updated. For 
instance, to interact with new features on the website or to fix bugs in the existing code. The browser 
will check regularly whether the service worker script specified during the registration has changed. If 
the newly fetched script is byte-different to the already installed one, the new one is seen as an 
updated version and will be installed. The time interval in which the browser checks for an updated 
service worker can vary, depending on settings like the \textit{updateViaCache} parameter during the 
registration and the \textit{Cache-Control:~max-age} HTTP header. 

It can be set to check on every navigation event on the registered scope, but it can also be 
postponed to an interval of maximal 24 hours. This means, the browser will swap an already running 
worker for a new version after a day, in case one is available. If no new service worker is available the 
service worker will stay within the browser as long as the user does not deleted it. The upper limit of 24 
hours is meant as a security measure, to prevent that a bug in a service worker could make the website 
unusable for the user. In this case, a repaired version of the service worker gets 
installed on the user's computer no later than a day. Once the old service worker is replaced, the 
updated one will start to receive all \textit{fetch} events for the registered scope. It will behave like a 
newly installed one as described before.

\begin{figure*}[t]
	\centerline{\includegraphics[width=0.8\textwidth]{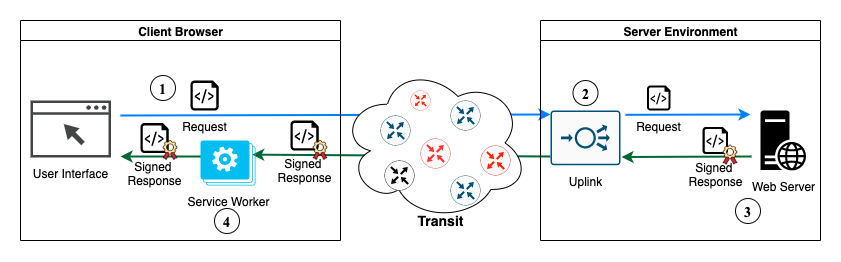}}
	\caption{Integrity validation with service workers.}
	\label{fig_sign_and_verify}
\end{figure*}
\section{Integrity Validation}\label{web_application_integrity_validation}
The capabilities of services workers are not limited to answer failed requests in the case of a network 
outage. They can in principle intercept any kind of request and answer them with an arbitrary 
response created by the service worker. They can also initiate their own requests and do all kinds of 
processing, as service workers are just regular instances of independently running JavaScript code.

Figure \ref{fig_sign_and_verify} shows an overview of our approach. Service workers are used to
detect manipulations of received web content and to seize appropriate measures once a manipulation 
is detected. We achieve this by integrating digital signatures into the header of all HTTP responses 
sent by 
the TLS endpoint. These signatures can be used by the service worker to verify the integrity of 
response bodies (e.g., containing JavaScript, HTML or other content).

The first step \circled{1} in Figure \ref{fig_sign_and_verify} represents a HTTPS request by a client 
browser to a specific domain. As shown in Figure \ref{fig_sign_and_verify} the request is sent over an 
unknown and untrusted network to the web applications uplink \circled{2}. Such an uplink can for 
instance be a load balancer or web application firewall which then forwards the request to a specific 
web server.

Depending on the network infrastructure either the web server \circled{3} or the uplink will 
then add a signature to the HTTP header, which is later used within the service worker for the integrity 
validation. The signature is generated with a private key over the HTTPS response body. Appending the 
signatures happens ideally on an upstream system which is less exposed to attacks than the web 
server. If an upstream system is not an option, the signature can also be added on the web server 
itself. The proposed separation is based on the requirement that this system must have access to the 
private key, which should be kept secret.

After the signature is appended to the response, it is sent further upstream where it is routed through 
transit zones (for example the internet) until it eventually reaches the browser of the requesting client. 
The service worker \circled{4} then intercepts all incoming requests and checks if the HTTP header 
contains a valid signature. The service worker can check the integrity of the web content by using the 
public key of the web server. In our example the public key is enrolled together with the service worker. 
In case of a valid signature the service worker forwards the response to the browsers user interface. In 
case of an invalid signature, domain specific steps can be taken. Such steps are further discussed in 
Section \ref{incident_handling}.

As soon as the service worker starts working, we can assume that all delivered web content was not 
tampered by an active man-in-the-middle. Moreover, to prevent a passive man-in-the-middle we can 
encrypt all sensitive web content in JavaScript. To put it another way, the service worker ensures that 
our JavaScript code is unchanged and JavaScript encryption protects against eavesdropping. In 
conclusion, active and passive man-in-the-middle attacks are no longer feasible, under the assumption 
that we can securely deliver a service worker.

\section{Implementation Example}\label{service_worker_integration}
As proof of concept, an example implementation of the described concept was developed. Within this 
section we discuss details and problems of the example implementation. The 
example will focus on how the signing process can be implemented with standard cryptographic 
functions and which measures can be taken in case of invalid signatures. Furthermore, in Section 
\ref{Replay_Attack_Protection} possible replay attack protection mechanism are shown.

Within the \textit{install} event the public-keys of the server are loaded into the service worker. As the 
install event is only executed once, it pins the authentication mechanism to a certain set of keys that 
remain constant for the lifetime of the service worker. This is done for simplicity to explain the idea 
behind the integrity validation. Other PKI solutions can be established. After all is set up, the waiting 
stage is skipped in order to replace a possibly active predecessor. When the \textit{activate} event is 
triggered, the scope of the website gets immediately claimed, so upcoming requests will flow through 
the service worker.

As soon as a \textit{fetch} event is caught, the essential part of the our solution takes place: the 
integrity validation. For this operation, the original requests from the client browser are forwarded to 
the web 
server and its response is awaited. In order to enable the service worker to detect data manipulations, 
we utilize regular digital signatures (e.g., ECDSA) and embed them in every HTTP response header from 
the web server. The service worker will validate all responses by generating the hash digest (SHA-256) 
of the received response body and check if it matches with the cryptographic signature in the 
header. Only if this check succeeds, will it forward the response to the regular part of the browser, 
which will then process and display this response in the DOM. If the signature is invalid or not present, 
the response will be discarded and an error message is returned to the users DOM instead.

\subsection{Incident Handling}\label{incident_handling}
Once the service worker receives content with an invalid signature, a way to deal with it is to 
discard the received response. This however will help neither the user nor the service provider to 
detect an ongoing attack. For the user it will just look as if they have a connection problem to the 
server and the service provider will not know that something went wrong. It is up to the service worker 
developer to decide which steps are taken when a invalid signature is found. 

For our example we implemented the following three steps, which allow the user and the service worker 
to get informed:

\textit{Client Session Termination:} If an incident is detected by the service worker, it can 
terminate the user's session on the client side. This can be done by deleting or overwriting a 
session cookie that keeps the user logged in. 

\textit{Client Visual Warning:} A service worker can inform the user about the reason why 
the session was terminated. In any case such an incident is not something the user should have to 
deal with solely by themselves. A visual warning can include the telephone number or e-mail 
address of the service provider for incident reporting.

\textit{Incident Reporting:} The service provider should be notified. If the provider can detect 
an accumulation of incidents they ideally can take targeted actions against the presumed attack. A 
question that arises in this scenario is the following: If the service worker detects manipulations 
despite an allegedly secure TLS connection, how can the service provider get informed about this 
without having the attacker just dropping this alert on the insecure HTTPS channel?

This depends on the location of the attacker. If they control all network traffic from the client to the 
Internet, one can not prevented them from filtering such an alert message. Nevertheless, if they just 
control a single connection (e.g., if they hijacked the web server or an upstream system in the 
data center) the service worker can try to send an alert message to another server with a different 
domain name (or multiple in various locations). It's assumed unlikely that an attacker can intercept 
TLS traffic from more than one certificate authority.

\subsection{Validating the Service Worker}\label{validating_the_service_worker_itself}
If a service worker is installed for a certain domain (or a scope to be precise), it can intercept and 
validate all requests and responses related to this domain. The request for fetching the 
service worker itself is handled differently internally in the browser. Instead of yielding a \textit{fetch}
event, which can be intercepted by the service worker, a separate channel is used which is invisible to 
the running service worker. This means, that if an attacker, holding a man-in-the-middle position, 
delivers a compromised service worker to the browser (e.g., one that just deems all unsigned content 
as valid), the running service worker is not able to prevent the modified service worker from getting 
installed. Nevertheless, as the browser checks for an updated service worker only when the rest of the 
webpage is loaded, it is possible to fetch a copy of the service worker through the genuine active 
service worker before the update within the browser. This means that even if an attacker is able to 
modify 
the service worker itself, a manipulation is detected and can invoke the incident handling routine. 
Service workers cannot be prevented from updating, but even in this case an attack can still be 
detected and incidents routines can be executed.

Another key thing to remember is that a service worker stays within the browser as long as there is no 
update or the user clears the browser cache. This means that a service worker can actively protect all 
the web content over months or even years. Additionally, several service 
workers can be active in parallel which allows to take incident handling actions based on specific 
scopes like sub-domains.

\subsection{Replay Attack Protection}\label{Replay_Attack_Protection}
With the presented approach an attacker is not able to arbitrarily forge an HTTP response that 
will be accepted as valid by the service worker. Nevertheless, without taking additional measures, an 
attacker can collect and record valid responses and just send them to the user. Depending on the 
application, this may trick a user into the execution of unintended actions when presented with old, 
replayed dynamic content. For static content this can mean that an attacker can send vulnerable 
versions of them, which have already been fixed by the developers (e.g., a vulnerable processing 
routine in JavaScript or an HTML with embedded elements, pointing to malicious content).

To solve this problem for dynamic content, a timestamp which is also sent in the HTTP header, gets 
pre- or appended to the response body, before hashing it for the digital signature. Depending on the 
nature of the web application all content older than a certain number of seconds can be discarded, 
even when bearing an otherwise valid signature.
Although modern signature algorithms should be resilient against Chosen Message Attacks (CMAs), 
this approach provides additional protection. As the message part chosen by the server will only 
change every second in a common timestamp, it is possible to concatenate some bytes of random data 
to the timestamp to make it unique and therefore prevent CMAs.

The situation for static content is a bit different: As static content may be cached and does not get 
signed freshly every time it is sent out, a time-based validation is impractical. Instead of a 
timestamp it is possible to include for example a build version number into the digital signature. The 
service worker can then keep track of the highest version number it has registered for static content 
yet, and discard all responses which arrive with a lower version number than that. This would be an 
additional security measure that could be configured if necessary.

\subsection{Content Delivery Networks}\label{cdns}
For scalability reasons a service provider might want to make use of third party content delivery 
networks to enhance the performance of the web application. This is possible with service worker and 
can be done in various ways, depending on the type of CDN.

CDNs that act as \textbf{dynamic proxy} and cache will work out of the box. They will just forward all 
content with the appropriate headers and replay until the \textit{cache-control: max-age} 
runs out. In case of key changes, a versioning scheme in the URLs will prevent clients from requesting 
content signed with expired keys. 

CDNs that deliver only \textbf{predeployed} static content are not able to deliver the 
appropriate HTTP headers for our approach. This is especially true for publicly hosted multi-purpose 
content like common JavaScript libraries. In this case it is possible to use subresource 
integrity~\cite{subresource_integrity, fetch_api_specification}. Where subresource integrity is not an 
option, the service provider can preload the service worker with a list of static file URLs and their 
appropriate signatures.
\vspace{-1ex}
\newpage
\section{Evaluation}\label{evaluation}
To evaluate whether our approach works as expected, we implemented a prototype system and 
performed a series of test covering the attack scenarios described in Section \ref{attack_scenarios}. 
Our test setup consisted of three components. The first component was a client with Google Chrome 
75 and Mozilla Firefox 67. These two browsers were used to send 10'000 requests per test to the 
server.
The second component was a web server with a node based web page with static (CSS, HTML, image 
assets) and dynamic content (JavaScripts). The third component was a proxy that could intercept and 
manipulate the requests and responses from the client and server. 

In order to simulate man-in-the-middle attacks with a rouge certificate, we installed a self-signed CA 
certificate on the client and server, issued and installed a corresponding server certificate on the 
server, and made the server certificate and the corresponding private key available to the proxy. 
To simulate attacks, the proxy was configured to intercept and modify server responses at defined 
rates from 0\% to 100\% by applying random sampling. 

The test for a given rate consisted of sending 10'000 requests per browser to the server and was 
considered successful if the service workers could identify all manipulated files without false positives.

\section{Results}\label{results}
These are the  results of the evaluation with regard to the three attack scenarios described in Section 
3: 

\textbf{Scenario (A):} No tests were carried out for this scenario as our approach cannot defend 
against man-in-the-middle attacks when the client has never visited the web page before. 
In this case, the service worker is not registered to the scope yet and no trust relationship is 
established. This is a limitation of all TOFU-based approaches.

\textbf{Scenario (B):} In this scenario, the client has visited the domain once to install the service 
worker. For all of the tests we set the interception rates between 1\% and 100\%, with increasing 
steps of one percent: the service worker was able to detect all of the manipulated files with no false 
positives. 

\textbf{Scenario (C):} In this scenario, the client had already an active TLS session to the web server 
after installing the service worker on its initial visit to the web server. As with scenario (B), for all of the 
tests we set the interception rates between 1\% and 100\%, with increasing steps of one percent: the 
service worker was able to detect all of the manipulated files with no false positives.

\section{Limitations}\label{discussion}
Within this section we discuss some of the mentioned limitations as well as future work of our 
approach.

\subsection{Trust On First Use}\label{tofu_discussion}
A service worker needs first to be installed before it can actively secure the 
communication. In other words, no attacks can be mitigated before a 
benign service worker has been installed in the client`s browser. When 
providing a website with a service requiring high security standards (e.g., 
e-banking or sensitive document management), we can assume that most 
users are regular users. Therefore chances that most users will have a 
genuine service worker installed over time and can profit from the enhanced 
protection are high. However, compared to other techniques a TOFU 
approach may not be an appropriate security standard for all use-cases and 
does not mitigate all possible attack vectors. The TOFU problem in 
enterprise environments may be avoided by supplying the 
client with the secrets over a secure channel.

\subsection{Key Management}\label{future_work_key_management}
Another attack vector is stealing the server's private key. In this case an attacker could bypass our 
security approach and sign malicious content with the stolen key. In order to mitigate the time 
window in which a stolen private key can be used, it is recommended to exchange the key material on a 
regular basis. This means that service workers must be able to handle these key updates. 

Our approach so far only mentions the use of one key pair. Using only one key pair is problematic 
since already installed service workers do not get the new key material until they are updated. So when 
the server starts signing the content with a new key all active service workers could no longer validate 
the integrity of the content, until the new public key is distributed to them. To prevent this drawback a 
key rotation or other PKI solutions can be used. 

Instead of placing one public key within the service worker, two or more public keys can be placed. This 
means that the service worker validates the signatures against a set of public keys, allowing the server 
to change the signing key on a regular basis without interrupting active service workers. The server 
just needs to rotate to the next private key and can replace deprecated public keys in the patched 
service worker (see Section~\ref{future_work_key_management}).

\subsection{Processing cost on the server}\label{future_work_performance}
The further costs to add another security layer to the TLS traffic consist of an additional hash- and 
sign-operation. Comprehensive performance tests with real traffic have yet to be run. A recent 
benchmark~\cite{zochowski2019benchmark} conducted using OpenSSL 1.1.1 provides an 
approximation to the additional costs on top of TLS traffic arising in a production environment: The 
extra time needed to secure a message of 1 KB was measured to be 3.78 ns to compute the hash 
(SHA-256) and 0.44 $\mu$s to sign it (ECDSA P-256) using a certain reference 
hardware\footnote{AWS t2.2xlarge instances with 8 vCPUs from an E5–2686 v4 CPU at 2.30GHz 
	running Ubuntu 16.04 with kernel 4.4.0}.

The results presented in Section~\ref{results} can be further improved by including the logic for adding 
the integrity validation information to the WAF solutions. We expect this measure to also reduce 
processing time.
On the client side, preliminary tests in our testing environment show no subjective noticeable delay in 
loading or using web applications augmented with our approach\footnote{Using browsers based on  
	Chromium 75 and Firefox 67.}. This is because all of the critical routines doing the 
heavy lifting are built-in by modern browsers and run at native speed (i.e., the Fetch 
API~\cite{fetch_api_specification} and the Web Crypto API~\cite{web_cryptography_api}). In the future, 
we will perform broader and more systematic tests in real-life scenarios to get realistic numbers on the 
introduced overhead.

\section{Conclusion}\label{conclusion}
We illustrated challenges on securing web applications with TLS and additional client-side security 
layers written in JavaScript. We introduced service workers and presented a 
novel approach to detect manipulated web application content utilizing 
them. We demonstrated how this approach 
protects web application users from passive and active man-in-the-middle 
attacks. Furthermore, we introduced a service worker based signing 
scheme, which safeguards users against malicious content coming from a 
compromised TLS connection. We built a proof of concept implementation 
of this scheme and evaluated our approach against an assumed attack 
scenario.

It was described how service worker can be used for an additional security measurement in order to 
protect web content delivery. The approach described in Section 
\ref{web_application_integrity_validation}, uses public key cryptography 
and a custom HTTP headerto guarantee the integrity of JavaScript files 
under some circumstances. Depending on the attack scenario one wants to 
prevent, service workers are a legit method to detect and defend against 
man-in-the-middle attacks. Despite the fact that service workers rely on a 
\textit{trust on first use} approach they can be an additional line of 
defence.

As shown service worker in combination with other security standard like 
CT can be especially useful when an attacker manages to issue rogue 
certificates. In such a case attackers would not only have 
to manage to issue a rogue certificate unnoticed, but also to replace all 
service workers on different scopes without breaking the web-applications, 
which can not go undetected as described in Section 
\ref{validating_the_service_worker_itself}.

All things considered, the approach shows that once a service worker is working on a domain 
it can be used to detect and mitigate ongoing man-in-the-middle attacks. Furthermore, the integrity 
validation of web content files with service workers allow to implement an additional layer of encryption 
within JavaScript. This encryption layer allows to defend against eavesdropping attacks, which gives 
the service provider and users a new way of securing their communication. 

The concept of using service workers as additional security measure to enhance TLS is not only limited 
to man-in-the-middle attacks. Another key idea behind using service workers is the detection of 
hacked web servers which delivery modified content. As mentioned in Section 
\ref{web_application_integrity_validation} service workers are used for adding a HTTP header with a 
signature to the web content. The system is considered secure as long as the private key is kept 
secret. This concept could further be enhanced with signing static (HTML, 
CSS, images etc.) and dynamic (JavaScript) content with separate keys. Such 
a separation could be done 
during the build process of a web application. 
In the scenario that the private key of the web server is stolen it would only 
allow an attacker to change parts (dynamic content) of the web application 
and a service worker could detect if the attacker attempts to change the 
static content.

Overall service workers can be used as an additional security measurement 
to prevent man-in-the-middle attacks but fail to fully secure a client from 
all possible attacks, due to the browsers API limitations when it comes to 
handling service worker updates and a TOFU problematic to deliver the 
service worker. 

\bibliographystyle{ACM-Reference-Format}
\bibliography{serviceWorker}


\begin{thebibliography}{18}


\ifx \showCODEN    \undefined \def \showCODEN     #1{\unskip}     \fi
\ifx \showDOI      \undefined \def \showDOI       #1{#1}\fi
\ifx \showISBNx    \undefined \def \showISBNx     #1{\unskip}     \fi
\ifx \showISBNxiii \undefined \def \showISBNxiii  #1{\unskip}     \fi
\ifx \showISSN     \undefined \def \showISSN      #1{\unskip}     \fi
\ifx \showLCCN     \undefined \def \showLCCN      #1{\unskip}     \fi
\ifx \shownote     \undefined \def \shownote      #1{#1}          \fi
\ifx \showarticletitle \undefined \def \showarticletitle #1{#1}   \fi
\ifx \showURL      \undefined \def \showURL       {\relax}        \fi
\providecommand\bibfield[2]{#2}
\providecommand\bibinfo[2]{#2}
\providecommand\natexlab[1]{#1}
\providecommand\showeprint[2][]{arXiv:#2}

\bibitem[\protect\citeauthoryear{??}{moz}{[n. d.]}]%
        {mozilla_ca_certificate_list}
 \bibinfo{year}{[n. d.]}\natexlab{}.
\newblock \bibinfo{title}{Mozilla Included CA Certificate List}.
\newblock
\newblock
\urldef\tempurl%
\url{https://wiki.mozilla.org/CA/Included_Certificates}
\showURL{%
\tempurl}
\newblock
\shownote{Accessed: 2019-06-14.}


\bibitem[\protect\citeauthoryear{??}{can}{[n. d.]}]%
        {caniuse_service_workers}
 \bibinfo{year}{[n. d.]}\natexlab{}.
\newblock \bibinfo{title}{Service Workers}.
\newblock
\newblock
\urldef\tempurl%
\url{https://caniuse.com/#feat=serviceworkers}
\showURL{%
\tempurl}
\newblock
\shownote{Accessed: 2019-06-14.}


\bibitem[\protect\citeauthoryear{??}{sup}{2015}]%
        {superfish}
 \bibinfo{year}{2015}\natexlab{}.
\newblock \bibinfo{title}{SuperFish Vulnerability}.
\newblock
\newblock
\urldef\tempurl%
\url{https://support.lenovo.com/ch/en/product_security/superfish}
\showURL{%
\tempurl}
\newblock
\shownote{Accessed: 2019-06-14.}


\bibitem[\protect\citeauthoryear{Akhawe, Braun, Marier, and Weinberger}{Akhawe
  et~al\mbox{.}}{2016}]%
        {subresource_integrity}
\bibfield{author}{\bibinfo{person}{Devdatta Akhawe}, \bibinfo{person}{Frederik
  Braun}, \bibinfo{person}{François Marier}, {and} \bibinfo{person}{Joel
  Weinberger}.} \bibinfo{year}{2016}\natexlab{}.
\newblock \bibinfo{title}{Subresource Integrity}.
\newblock
\newblock
\urldef\tempurl%
\url{https://w3c.github.io/webappsec-subresource-integrity/}
\showURL{%
\tempurl}
\newblock
\shownote{Accessed: 2019-09-02.}


\bibitem[\protect\citeauthoryear{B.~Laurie}{B.~Laurie}{2013}]%
        {certificateTransparency}
\bibfield{author}{\bibinfo{person}{E.~Kasper B.~Laurie, A.~Langley}.}
  \bibinfo{year}{2013}\natexlab{}.
\newblock \bibinfo{booktitle}{\emph{{Certificate Transparency}}}.
\newblock \bibinfo{type}{{RFC}} 6962. \bibinfo{institution}{{Internet
  Engineering Task Force (IETF)}}. \bibinfo{pages}{1--27} pages.
\newblock
\showISSN{2070-1721}
\urldef\tempurl%
\url{https://www.rfc-editor.org/rfc/pdfrfc/rfc6962.txt.pdf}
\showURL{%
\tempurl}


\bibitem[\protect\citeauthoryear{Berkowsky and Hayajneh}{Berkowsky and
  Hayajneh}{2017}]%
        {berkowsky2017security}
\bibfield{author}{\bibinfo{person}{Jake~A Berkowsky} {and}
  \bibinfo{person}{Thaier Hayajneh}.} \bibinfo{year}{2017}\natexlab{}.
\newblock \showarticletitle{Security issues with certificate authorities}. In
  \bibinfo{booktitle}{\emph{Ubiquitous Computing, Electronics and Mobile
  Communication Conference (UEMCON), 2017 IEEE 8th Annual}}. IEEE,
  \bibinfo{pages}{449--455}.
\newblock


\bibitem[\protect\citeauthoryear{C.~Evans}{C.~Evans}{2015}]%
        {httpKeyPinningPaper}
\bibfield{author}{\bibinfo{person}{R.~Sleevi C.~Evans, C.~Palmer}.}
  \bibinfo{year}{2015}\natexlab{}.
\newblock \bibinfo{booktitle}{\emph{{Public Key Pinning Extension for HTTP}}}.
\newblock \bibinfo{type}{{RFC}} 7469. \bibinfo{institution}{{Internet
  Engineering Task Force (IETF)}}. \bibinfo{pages}{1--28} pages.
\newblock
\showISSN{2070-1721}
\urldef\tempurl%
\url{https://www.rfc-editor.org/rfc/pdfrfc/rfc7469.txt.pdf}
\showURL{%
\tempurl}


\bibitem[\protect\citeauthoryear{Gallagher}{Gallagher}{2013}]%
        {nsa_flying_pig}
\bibfield{author}{\bibinfo{person}{Ryan Gallagher}.}
  \bibinfo{year}{2013}\natexlab{}.
\newblock \bibinfo{title}{New Snowden Documents Show NSA Deemed Google Networks
  a Target}.
\newblock
\newblock
\urldef\tempurl%
\url{https://slate.com/technology/2013/09/shifting-shadow-stormbrew-flying-pig-new-snowden-documents-show-nsa-deemed-google-networks-a-target.html}
\showURL{%
\tempurl}
\newblock
\shownote{Accessed: 2019-06-14.}


\bibitem[\protect\citeauthoryear{Hickson, Pieters, van Kesteren, Jägenstedt,
  and Denicola}{Hickson et~al\mbox{.}}{[n. d.]}]%
        {html_living_standard}
\bibfield{author}{\bibinfo{person}{Ian Hickson}, \bibinfo{person}{Simon
  Pieters}, \bibinfo{person}{Anne van Kesteren}, \bibinfo{person}{Philip
  Jägenstedt}, {and} \bibinfo{person}{Domenic Denicola}.} \bibinfo{year}{[n.
  d.]}\natexlab{}.
\newblock \bibinfo{title}{HTML Living Standard}.
\newblock
\newblock
\urldef\tempurl%
\url{https://html.spec.whatwg.org}
\showURL{%
\tempurl}
\newblock
\shownote{Accessed: 2019-06-14.}


\bibitem[\protect\citeauthoryear{Ormandy}{Ormandy}{2016}]%
        {kaspersky_mim}
\bibfield{author}{\bibinfo{person}{Tavis Ormandy}.}
  \bibinfo{year}{2016}\natexlab{}.
\newblock \bibinfo{title}{Kaspersky: SSL interception differentiates
  certificates with a 32bit hash}.
\newblock
\newblock
\urldef\tempurl%
\url{https://bugs.chromium.org/p/project-zero/issues/detail?id=978}
\showURL{%
\tempurl}
\newblock
\shownote{Accessed: 2019-06-14.}


\bibitem[\protect\citeauthoryear{P.~Hoffman}{P.~Hoffman}{2012}]%
        {dane}
\bibfield{author}{\bibinfo{person}{J.~Schlyter P.~Hoffman}.}
  \bibinfo{year}{2012}\natexlab{}.
\newblock \bibinfo{booktitle}{\emph{{The DNS-Based Authentication of Named
  Entities (DANE) Transport Layer Security (TLS) Protocol: TLSA}}}.
\newblock \bibinfo{type}{{RFC}} 6698. \bibinfo{institution}{{Internet
  Engineering Task Force (IETF)}}. \bibinfo{pages}{1--35} pages.
\newblock
\showISSN{2070-1721}
\urldef\tempurl%
\url{http://buildbot.tools.ietf.org/html/rfc6698}
\showURL{%
\tempurl}


\bibitem[\protect\citeauthoryear{Palmer}{Palmer}{2017}]%
        {google_remove_hpkp}
\bibfield{author}{\bibinfo{person}{Chris Palmer}.}
  \bibinfo{year}{2017}\natexlab{}.
\newblock \bibinfo{title}{Intent To Deprecate And Remove: Public Key Pinning}.
\newblock
\newblock
\urldef\tempurl%
\url{https://groups.google.com/a/chromium.org/forum/#!msg/blink-dev/he9tr7p3rZ8/eNMwKPmUBAAJ}
\showURL{%
\tempurl}
\newblock
\shownote{Accessed: 2019-06-14.}


\bibitem[\protect\citeauthoryear{Prins and Cybercrime}{Prins and
  Cybercrime}{2011}]%
        {prins2011diginotar}
\bibfield{author}{\bibinfo{person}{J~Ronald Prins} {and}
  \bibinfo{person}{Business~Unit Cybercrime}.} \bibinfo{year}{2011}\natexlab{}.
\newblock \showarticletitle{Diginotar certificate authority breach’operation
  black tulip’}.
\newblock \bibinfo{journal}{\emph{Fox-IT, November}} (\bibinfo{year}{2011}).
\newblock


\bibitem[\protect\citeauthoryear{Russell, Jungkee, Archibald, and
  Kruisselbrink}{Russell et~al\mbox{.}}{2017}]%
        {service_workers_editors_draft}
\bibfield{author}{\bibinfo{person}{A. Russell}, \bibinfo{person}{S. Jungkee},
  \bibinfo{person}{J. Archibald}, {and} \bibinfo{person}{M. Kruisselbrink}.}
  \bibinfo{year}{2017}\natexlab{}.
\newblock \bibinfo{title}{Service Workers 1}.
\newblock
\newblock
\urldef\tempurl%
\url{https://www.w3.org/TR/2017/WD-service-workers-1-20171102/}
\showURL{%
\tempurl}
\newblock
\shownote{Accessed: 2019-06-14.}


\bibitem[\protect\citeauthoryear{Slootweg}{Slootweg}{[n. d.]}]%
        {certificate_authority_incidents}
\bibfield{author}{\bibinfo{person}{Sven Slootweg}.} \bibinfo{year}{[n.
  d.]}\natexlab{}.
\newblock \bibinfo{title}{Certificate Authority Incidents}.
\newblock
\newblock
\urldef\tempurl%
\url{https://git.cryto.net/joepie91/ca-incidents}
\showURL{%
\tempurl}
\newblock
\shownote{Accessed: 2019-06-14.}


\bibitem[\protect\citeauthoryear{van Kesteren}{van Kesteren}{[n. d.]}]%
        {fetch_api_specification}
\bibfield{author}{\bibinfo{person}{Anne van Kesteren}.} \bibinfo{year}{[n.
  d.]}\natexlab{}.
\newblock \bibinfo{title}{Fetch API}.
\newblock
\newblock
\urldef\tempurl%
\url{https://fetch.spec.whatwg.org/}
\showURL{%
\tempurl}
\newblock
\shownote{Accessed: 2019-06-14.}


\bibitem[\protect\citeauthoryear{Watson}{Watson}{2017}]%
        {web_cryptography_api}
\bibfield{author}{\bibinfo{person}{Mark Watson}.}
  \bibinfo{year}{2017}\natexlab{}.
\newblock \bibinfo{title}{Web Cryptography API}.
\newblock
\newblock
\urldef\tempurl%
\url{https://www.w3.org/TR/2017/REC-WebCryptoAPI-20170126/}
\showURL{%
\tempurl}
\newblock
\shownote{Accessed: 2019-06-14.}


\bibitem[\protect\citeauthoryear{Zochowski, Hua, and Wang}{Zochowski
  et~al\mbox{.}}{2019}]%
        {zochowski2019benchmark}
\bibfield{author}{\bibinfo{person}{Michael Zochowski}, \bibinfo{person}{Carl
  Hua}, {and} \bibinfo{person}{Peng Wang}.} \bibinfo{year}{2019}\natexlab{}.
\newblock \bibinfo{title}{Benchmarking Hash and Signature Algorithms}.
\newblock
\newblock
\urldef\tempurl%
\url{https://medium.com/logos-network/benchmarking-hash-and-signature-algorithms-6079735ce05}
\showURL{%
\tempurl}
\newblock
\shownote{Accessed: 2019-09-02.}


\end{thebibliography}

\end{document}